\documentstyle[aps,twocolumn]{revtex}

\begin{document}
\draft
\wideabs{
\title{Magnetic properties of superconducting
multifilamentary tapes in perpendicular field. II: Horizontal and
matrix arrays}
\author{Enric Pardo$^{\rm a}$, Carles
Navau$^{\rm a,b}$, and Alvaro Sanchez$^{\rm a}$}
\address{$^{\rm a}$ Grup d'Electromagnetisme,
Departament de F\'\i sica, Universitat Aut\`onoma Barcelona \\
08193 Bellaterra (Barcelona), Catalonia, Spain\\
$^{\rm b}$ Escola Universit\`aria Salesiana de Sarri\`a, Passeig
Sant Joan Bosco 74, 08017 Barcelona, Catalonia, Spain}
\maketitle
\begin{abstract}

Current and field profiles, and magnetization and ac losses are
calculated for arrays of infinitely long superconducting tapes in the
critical state
in a perpendicularly applied magnetic field. The tapes are arranged
both horizontally and in a matrix configuration, which is the
geometry found in many actual high-$T_c$ superconducting tapes. 
The finite
thickness of the tapes and the effects
of demagnetizing fields are considered. Systematic results for the
magnetization and ac losses of the tapes are obtained as function of
the geometry and separation of the constituent tapes. Results allow
to understand some unexplained features observed in experiments, as
well as to propose some future directions.
\end{abstract}
}

\section{Introduction}

High temperature superconducting (HTSC) cables have a large potential
for many applications where a very high current intensities are
needed, such as power transmission cables, magnets, superconducting
magnetic energy storage systems (SMES), transformers, and motors
\cite{appl1,appl3}. 
In particular, silver sheathed ${\rm Bi_2Sr_2Ca_2Cu_3O_{10}}$
(Ag/Bi-2223) tape conductors showed to be the HTSC most used for
practical devices, due to the good superconductor material quality
and the feasibility to make kilometer long cables. Many of the HTSC
cables applications work under AC conditions, like power transmission
cables, transformers and motors. An important problem for the
superconducting power devices operating at AC intensities is caused
by their power losses \cite{ACloss}, which must be reduced as low as
possible to justify the expenses of the superconducting material and
the cryogenic system. We can distinguish between self field AC
losses,
that is, the power losses due to transport current inside each
conductor, and the magnetic AC losses due to a magnetic field
external to the conductor, which we deal with in this work. The
latter kind of losses are important for devices where a high magnetic
field is present, like magnets and transformers.

Magnetic AC losses critically depend on the superconductor wire
geometry \cite{Wilson,Carr}.
As it was pointed out in \cite{Wilson,Carr,Oota}, dividing the
superconductor wire into filaments reduces the magnetic losses.
Moreover, it is known that dividing superconducting wires into
filaments and immersing them into a conducting matrix makes the wire
more reliable under quenching \cite{Wilson,Carr}. In addition, it is
shown that for Ag/Bi-2223 tapes, the superconducting
properties improves when the superconducting region is divided into
filaments with a high aspect ratio \cite{qual1,qual2}. Then, this is
the HTSC wire geometry most often met in practice. 

The magnetic AC losses in multifilamentary tapes have their origin in
mainly three mechanisms. They are the eddy currents in the conducting
sheath, the magnetic hysteresis arising from the flux pinning in the
superconductor, and the inter-filament currents (also known as
coupling currents) that flow across the conducting matrix
\cite{Wilson,Carr}.
Although it is somehow understood how to reduce the eddy and coupling
currents losses \cite{Fukumoto95,muller}, important work remains to
be done
concerning the
hysteresis losses. Many experimental works showed that the
hysteresis losses depend strongly on the orientation of the external
AC field \cite{OomenAPL,Wolfbrandt,Chiba}. It is shown that the
hysteresis losses under an applied field $H_{\rm a}$ perpendicular to
the wide face of the HTSC tapes are more than one order of magnitude
higher than if $H_{\rm a}$ is either parallel to the wide face or in
the transport direction. 

Although the losses for $H_{\rm a}$ parallel
to the tape wide face or in the transport direction are theoretically
well described by the Bean's model for a slab
\cite{Bean,Fukumoto95,Fukumoto95b}, there is not any theoretical
model that satisfactorily describe the losses of multifilamentary
tapes under perpendicular $H_{\rm a}$ \cite{GomAC_00,GomAC_02}.
The only theories so far describe partially the
magnetic properties and hysteresis losses of multifilamentary tapes,
in cases
such as infinite $z$-stacks and $x$-arrays of infinitely thin strips
in the critical state model \cite{Mawatari}, realistic
multifilamentary tapes but considering complete shielding only
\cite{fabbricatore,Farinon}, and $z$-stacks of strips
\cite{tapesa,Tebano}. 

In the first paper of this series \cite{tapesa} we introduced a
general model for calculating the magnetic response of 
a finite thickness superconductor of infinite length within the
critical state model \cite{Bean} applied to a vertical stack of
infinitely long
superconducting strips. A case most often encountered in
real superconducting cables is that of an array of superconducting
filaments arranged in rows and columns. The purpose of this paper is
to numerically calculate and discuss the main magnetic properties of
superconducting multifilamentary tapes, such as field profiles,
magnetization curves, and magnetic AC
losses within the critical state model in a perpendicular applied
field. We consider the realistic
case that arrays have a finite number of filaments, each having
finite thickness. For all cases the external applied field $H_{\rm
a}$ is
considered to be uniform and perpendicular to the infinite dimension. 

The are two cases we study separately. In the first
one, current is restricted to return through the same filament. We
refer to this case as isolated filaments. This is the desired case
for AC magnetic losses reduction in real HTSC tapes
\cite{Wilson,Carr,Oota,GomAC_97}. The other case is when current can
go in one direction in a given filament and
return through another one. We
refer to the latter case as completely interconnected filaments. This
is the
limiting case of filaments with a high number of intergrowths
\cite{Volkozub99,Volkozub98,Everett98,Ashworth97,Glowaki99} or when
coupling currents through the conducting matrix are of the same
magnitude as the superconducting currents
\cite{Fukumoto95,GomAC_97,GomAC_98,Bobyl00}. As explained in these
references and below, the magnetic behavior for each
filament connection case is strongly different. Therefore, a detailed
study of ac losses in superconducting cables should include these two
cases. The strong difference in considering interconnected or
isolated strips can be realized in the current and field profiles
shown in Fig. \ref{f:Cpdp2} for horizontal arrays. 

Although in the present paper only calculations for filaments with a
high aspect ratio like those present in the actual tapes are shown,
the numerical method presented below has
been checked to be useful for any thickness-to-width ratio between
$0.001$ and $100$.

The present paper is structured as follows. In section II we present
the calculation model and its modifications from the original one in
\cite{tapesa}. Current and field profiles are calculated and
discussed in Section III. The results of magnetization and magnetic
AC losses are discussed in Section IV and V, respectively. Finally,
in Section VI we present the main conclusions of this work. The full
penetration field for $x$-arrays and $xz$ matrices can be
analytically calculated, being described in the Appendix.

\section{Model}

We assume the $x$-arrays and $xz$ matrices to be made up of identical
rectangular strips, which we consider infinitely long in the $y$
direction, as in \cite{tapesa}. The separation of the $xz$-matrix
rows is $h$ and the
separation of columns is $d$ (an $x$-array can be
considered as a matrix with a single row), as shown in Fig.
\ref{f:SkGn}. The strips have dimensions $2a$ and $2b$ in the
horizontal and vertical directions, respectively, and are divided
into
$2n_x \times 2n_z$ elements with cross-section $(\Delta x)(\Delta
z)$. We consider that, if present, the current density that flows
though each element is uniform.

As explained above, we discuss two different cases of filament
connection: when the filaments are all interconnected or when they
are
isolated to each other. To describe both cases we use mainly the same
model previously used for $z$-stacks \cite{tapesa} and for cylinders
\cite{cilI,cilII}, which is based on minimizing the magnetic energy
of current distribution within the critical state model. We name this
model as the minimum magnetic energy variation procedure, thereafter
referred to as MMEV. As
in \cite{tapesa}, we assume there is no equilibrium magnetization nor
field dependence of $J_c$. The method allows the calculation of
$J(x,z)$ in the initial magnetization process of a zero-field cooled
superconductor, from which the initial
magnetization curve, the complete magnetization loop and the AC
susceptibility can be easily calculated \cite{tapesa}. 

The numerical procedure explained below is valid for any
superconductor wire geometry as
long as it has $yz$ mirror symmetry. So, this model can be applied
to describe realistic multifilamentary tapes, and not only $x$-arrays
and $xz$ matrices.

We now discuss two features that are needed to apply the
MMEV procedure to a certain superconductor geometry. This will help
us to determine which modifications, if necessary, have to be done to
adapt the MMEV procedure to $x$-arrays and $xz$ matrices for 
the cases of interconnected and isolated strips.  

The first condition to apply the MMEV procedure is that one needs to
know the shape of the closed
current loops of the magnetically induced current for any applied
field value $H_{\rm a}$ (that is, one has to know what is the
returning
current element for a given one, so that both form a close current
loop).  
For cylinders the closed current loops
were simply circular \cite{cilI,cilII}, while for $z$-stacks
\cite{tapesa} they were made up by an infinite straight current in
the $y$ direction centered at $(x,z)$ and by another centered at
$(-x,z)$, which formed a closed circuit at infinity. 
For the latter case, the sign criterion for the current flowing in a
circuit is to take it positive when the current in the $x\ge 0$
region
follows the positive $y$ axis, and negative otherwise. 

Another feature that has to be taken into account in order to apply
the numerical method in Ref.
\cite{tapesa}, developed for $z$-stacks, to $x$-arrays and $xz$
matrices is the sign of the induced current. For simple
geometries such as rectangular strips and disks
\cite{ThStrEPL,ThStrPRB,ThickStr,Prigozhin,ClemSanchez}, elliptical
tapes
\cite{Prigozhin,ellipse,Bhagwat}, and $z$-stacks
\cite{Mawatari,tapesa}, current in the initial magnetization curve is
negative for all circuits, that is, current profiles are symmetrical
to the $yz$ plane. However, this feature it is not so
obvious for geometries with gaps in the horizontal direction like $x$
arrays and $xz$ matrices. 

As explained in the following subsections, the applicability of
the mentioned two features is different when considering
interconnected or isolated strips. So, the adaptation of the MMEV
procedure to the two different connection cases must be considered
separately.

\subsection{Interconnected strips}
\label{m_ins}

For the present case of $x$-arrays and $xz$-matrices with completely
interconnected strips the two features presented above are still
valid. This can be justified as follows.

First, the closed circuits to be used in the simulations are the same
as for $z$-stacks,
which are each pair of current elements centered at $(x,z)$ and
$(-x,z)$, Fig. \ref{f:SkGn}. This is so because of the mirror $yz$
symmetry of the system and the fact that the strips are
interconnected at infinity so that currents belonging to different
strips can be closed. 

Secondly, the fact that in the initial magnetization curve the
current is negative in all the circuits is also valid for the present
case. We arrive to this
conclusion after doing some preliminary numerical calculations, in
which we changed the original numerical
method letting the procedure to choose which sign in each circuit is
optimum to minimize the energy. After doing so, we saw that in the
initial magnetization curve and for a given $H_{\rm a}$, current is
the
same and negative for all circuits, except for very few circuits on
the final current profile due to numerical error. Notice that this
means that the current of the strips at the $x\ge 0$ region return to
those in the $x\le 0$ region, so that current return through
different strips for all circuits except for those centered at $x=0$.
This result is the expected one, because this situation is the one
that minimizes the most the energy, so it should be the chosen one
when there are no restrictions.

Then, we conclude that numerical method and formulae for $x$-arrays
and $xz$ matrices are the same as those previously used for
$z$-stacks \cite{tapesa} with the only modifications needed for
adapting the model to the new geometry.

\subsection{Isolated Strips}

The model used to describe current isolated strips must take into
account that all real current loops have to be closed inside each
strip, so that there has to be the same amount of current following
the
negative $y$ direction than the positive one inside each strip.
In addition, although the current distribution of the whole $x$-array
or $xz$-matrix have $yz$ mirror symmetry for the plane $x=0$, the
current distribution in the individual strips is not necessarily
symmetrical to
their $yz$ central plane, except for those centered at $x=0$.

Then, the features of the MMEV procedure
described above do not apply, so that we need to do significant
modifications to the original numerical procedure presented in
\cite{tapesa}.

The actual current loops in this case have the shape of two straight
lines within the same strip carrying opposite currents and closed at
infinity (solid lines in Fig. \ref{f:SkIs}). These straight currents
can be identified with the elements which the strips are divided in.
The main difficulty is to know which pairs of elements describe
closed current loops. 

To help solving this problem we notice that, thanks to the overall
mirror symmetry to the $yz$ plane at $x=0$, for any closed current
loop in a strip at the $x\ge 0$ zone, there is another current loop
set
symmetrically in the corresponding strip in the $x\le 0$ zone (Fig.
\ref{f:SkIs}). Furthermore, if we take as closed current loops the
pairs of
elements set symmetrically to the $yz$ plane (dashed lines in Fig.
\ref{f:SkIs}), the total current distribution is the same except at
the
ends, which do not modify the magnetic moment if we consider
the strips long enough.
Both systems of closed circuits have the same magnetic properties,
including magnetic energy and magnetic moment. Consequently, since
these symmetrical pairs of elements correspond to the closed loops
used for $z$-stacks in \cite{tapesa}, all the formulae presented
there are still applicable. 

Taking these symmetrical pairs of elements as closed loops for the
numerical procedure, as done in Sec. \ref{m_ins} for interconnected
strips, and the fact that current loops must close
inside each strip, the MMEV procedure for isolated strips becomes in
this case:

\begin{enumerate}
\item {For a given applied field $H_{\rm a}$, a given current
distribution,
and for each pair of strips set symmetrically to the $yz$ plane
(symmetrical pair of strips), there are found: i) the
loop where setting a negative current would reduce the
most the magnetic energy and ii) the loop where setting
a positive current would rise it the least. These loops are referred
to 
as a pair of loops.}

\item {The pair of loops that lower the most the magnetic
energy is selected among all those belonging to each symmetrical pair
of strips.}

\item {A current of the corresponding sign is set in the selected
loops.}

\item {This procedure is repeated until setting current in the most
energy-reducing pair of loops would increase the energy
instead of lowering it.}
\end{enumerate}

Notice that each pair of loops where current is set in the
simulations describes two
real closed current loops belonging to each strip that constitute the
symmetrical pair of strips.

\section{Current penetration and field profiles}

For the sake of clarity, we discuss separately the results
corresponding to the situation in which tapes are interconnected and
that in which they are isolated. 

\subsection{Interconnected strips}

We first discuss the current and field penetration profiles
calculated for an $x$-array composed of three filaments with
dimensions $b/a=0.1$. In Fig. \ref{f:FfInA} we show the current
profiles and the field lines corresponding to three $x$-arrays
with varying separation between the individual tapes. The applied
field in all cases is $0.2H_{\rm pen}$, being $H_{\rm pen}$ the
penetration field for the whole $x$-array (Appendix). The
common behavior observed is that currents are induced to try to
shield not only the superconductors (field is zero in the
current-free regions inside the superconductors) but also the space
between them. Actually, we find that there appears an overshielding
near the inner edge of the external strips (Fig. \ref{f:FfInA}), so
that the field there is opposite to the external field. This feature
has been previously predicted for rings in the critical state
\cite{BrandtRing} and for completely shielded toroids \cite{toroid}.

The general trends described above for the case of an $x$-array are
also
valid for the case of an $xz$-matrix. Actually, it is important to
remark that the general trends in current and field penetration and
the magnetic behavior of an $xz$-matrix
result from the composition of the properties of both the $x$-array
and the $z$ vertical stack that forms it. In Fig. \ref{f:FfInM} we
show the calculated current penetration profiles for an $xz$-matrix
made of nine strips ($3\times 3$), each with dimensions $b/a=0.1$
corresponding to an applied field of $0.2H_{\rm pen}$. We also plot 
the total (left figures) and self (right) magnetic field, that is,
the sum of the external magnetic field plus that created by the
superconducting currents and only the latter contribution,
respectively. The general trend of shielding the internal volume of
the region bounded by the superconductor, including gaps between
tapes, is also clearly seen. An interesting feature is that a very
satisfactory magnetic shielding is achieved for the three different
arrays, as illustrated from the fact that the self-field in the
central region has in all cases a constant value over a very large
region. However, this shielding is, for the values of the applied
field considered here, basically produced by the tapes
in the two outer vertical columns, which are largely penetrated by
currents. Only a little current is needed to flow in the upper and
bottom tapes of the inner column to create a fine adjustment of the
field in the central region.

\subsection{Isolated strips}

We now present the results calculated for the case that the
superconducting strips are isolated so that current has to go and
return always though the same filament. We start again with the
case of an $x$-array composed of three strips with dimensions
$b/a=0.1$. In Fig. \ref{f:FfIsA} we show the current profiles and the
field distribution calculated for three $x$-arrays with varying
separation between the individual tapes. The applied field is
$0.1H_{\rm pen}$. Again, all tapes have dimensions $b/a=0.1$. By
simple inspection, one can realize how important the differences are
with respect to the case of interconnected strips. In the present
case of isolated strips there is appreciable current penetration in
all the strips and not only in outer ones,
although the magnetic coupling between them makes the current
distribution in the outer strips different from the central one.
Another important effect to be remarked is that there is an important
flux compression in the space between the strips. Since all strips
tend to shield the magnetic field in their interiors, the field in
the air gap between each
pair of strips is stronger because fo the field
exclusion in both adjacent strips. Actually, field lines
are very dense not only in the gap between strips but also in a zone
in the strips nearest to the gap, where current penetrate an
important distance (this effect is particularly clear for the case of
the smallest separation). This compression effect was also found by
Mawatari for the case of $x$-arrays of very thin strips
\cite{Mawatari}, by Fabbricatore et
al for $x$-arrays, $xz$ matrices and realistic shapes of
multifilamentary tapes in the Meissner state
\cite{fabbricatore,Farinon}, and by Mikitik
and Brandt for a completely shielded double strip \cite{dcVBrandt}.

We can better compare the current and field profiles for the
interconnected and isolated cases by looking at figure \ref{f:Cpdp2},
where we plot current profiles for the $x$-array with
separation $d=0.2a$ for both interconnected and isolated cases. It
can be seen that for interconnected strips, current penetrates
earlier
(that is, for lower values of the applied field) in the outer tapes,
since currents flowing there create an important shielding not only
in each strip but in the whole space between them. On the other hand,
in the isolated strips case currents returning through the same strip
create a field compression in the channels (that include the gaps
and a portion of each strip near the gap), so that the amount of
current penetration is similar for the three strips. The current
distribution when the strips are close to each other is slightly
asymmetric with respect to the central plane of each strip, because
the field in the channels felt by
the inner sides of the outer strips has a different spatial
distribution than the homogeneous applied field felt in the outer
sides. We have found that this asymmetry increases
for thicker filaments, that is, higher $b/a$ (not shown). 

We now present some results for the $xz$-matrix array for the case of
isolated strips. As said above,
results for the matrix can be understood from the composition of the
effects of horizontal and vertical arrays. In Fig. \ref{f:FfIsM} we
show the calculated current penetration profiles for an $xz$-array
made of nine strips ($3\times 3$) with dimensions $b/a=0.1$
corresponding to an applied field of $0.1H_{\rm pen}$, together
with
the total (left figures) and self (right) magnetic field. The two
cases correspond to separations of $d/a=h/a=0.02$ and 2,
respectively. The effect
of flux compression along the vertical channels that include the gaps
and the surrounding regions is clearly seen in the case of the
smallest separation distance. 

For the case of $xz$ matrices it can be observed how field is
shielded
in
the vertical gaps between rows but it is enhanced in the horizontal
gaps between columns. Then, for isolated strips, magnetic interaction
between rows and columns have opposite effects. Furthermore, the
difference between the field in the vertical gaps and the applied
field $H_{\rm a}$ is much
higher than for the horizontal gaps, as can be seen in Fig.
\ref{f:FfIsM} for the $xz$-matrix with higher separation. This
implies that the magnetic coupling between strips in the horizontal
direction is lost at smaller distances than that in the vertical
direction.

\section{Magnetization}

\subsection{$x$-arrays}

We now analyze the results for the magnetization of the $x$-arrays.
In Fig. \ref{f:MAllA} we plot the calculated magnetization $M$ as
function of the applied field $H_{\rm a}$ for the 3 $x$-arrays of
Figs. \ref{f:FfInA} and \ref{f:FfIsA}. For each strip $b/a=0.1$, and
the separation distance between strips is, for each case, $d/a=0.02$,
0.2 and 2. The upper figure shows the results for the isolated strips
whereas the data in the bottom part is for interconnected
strips. 

The magnetization
for both isolated and interconnected strips shows important
differences,
arising from the different current penetration profiles studied in
the Section III. We first discuss the results for isolated
strips. It
can be seen that $M$ saturates at
smaller values than for the case of interconnected strips and that
this
saturation value is the
same for the different separations. The
results for largest separation, $d/a=2$, are not very
different from the results obtained from a single strip with
$b/a=0.1$, corresponding to the limit of complete magnetically
uncoupled strips,
which is also shown in the figure.  
An important result is that the initial slope of the
$M(H_{\rm a})$ curve $\chi_0$ increases (in absolute value) with
decreasing
separation. 
The reason
for this behavior can be traced back to the presence of the flux
compression effect discussed in Section IIIB, since a smaller
separation means thinner channels and a corresponding larger flux
compression. The enhancement of the initial slope can also be
explained
by the fact that the strips have to shield not only the external
applied field but also the field created by the other strips. This
enhancement of $\chi_0$ have already been predicted in similar
situations
\cite{Mawatari,fabbricatore,Farinon,Everett98}.

We have found that the initial slope
calculated with our approach is coincident, within a $4\%$
difference, with that calculated numerically by finite elements by
Fabbricatore et al \cite{fabbricatore} for the case of $5\times 5$
and $5\times 3$ filament matrices with complete shielding. The
initial slope has also been compared with other works
\cite{StrMeis,demstrI} for a single strip for a high range of $b/a$
($0.001\le b/a \le 100$), obtaining a difference smaller than a 1\%.

When comparing the results for isolated strip to the case of
interconnected ones, important differences appear. A first 
difference is that the saturation magnetization in the latter case
is not only larger in general with respect to the isolated case but
also
depends on the separation. The second difference is that the trend
found when decreasing the separation distance between tapes is
reversed: whereas for isolated strip decreasing separation distance
$d/a$ results in a larger (in absolute value) slope of the initial
magnetization, for interconnected strips the slope gets smaller with
decreasing separation. 

We explain the reasons for both differences as follows. The
difference behavior in the saturation magnetization arises from the
fact that this value corresponds to the magnetic moment per unit
volume when all the strips are fully penetrated. The magnetic
moment is proportional to the area threaded by the current loops,
which in interconnected case are not restricted to a single strip but
they can span from even one extreme of the array to the other.
Actually, the saturation magnetization $M_s$ can be
analytically calculated considering that, for isolated strips, at
saturation the $\pm J_c$ interface is close to a straight line, so
that $M_s$ is the same as for a single uncoupled strip, being
$M_s=1/2 J_c a$ \cite{Bean,ThStrEPL,ThStrPRB}. For the case of
interconnected strips, the current distribution at saturation is
${\bf J}=-J_c{\bf \hat y}$ for $x\ge 0$ and ${\bf J}=J_c{\bf \hat y}$
for $x\le 0$, so that $M_s$ can be calculated as
\begin{eqnarray}
M_s&=&{J_c\over 2}(2a+d)\left({n_{f,x}\over 2}\right)
\qquad\qquad\qquad\ \, (n_{f,x}\ {\rm even}) \\
M_s&=&{J_c\over 2n_{f,x}}\left[{ a+(2a+d){(n_{f,x}^2-1)\over 2}
}\right] \quad (n_{f,x}\ {\rm odd}),
\end{eqnarray}
where $n_{f,x}$ is the number of strips in the $x$ direction for
either an $x$-array or an $xz$-matrix.

As to
the initial slope of the $M(H_{\rm a})$ curve, in the case of
interconnected
tapes the flux compression effect discussed above does not exist so
the reason for the behavior of the initial slope of the $M(H_{\rm
a})$ curve must be a different one.
The governing effects now are the demagnetizing effects
arising from the large aspect ratio of the $x$-array taken as a
whole. The demagnetizing effects tend to enhance the initial slope
\cite{cilI,demstrII,cilMeis} when the sample aspect ratio increases.
Therefore, when the separation is small the array is behaving
similarly to a single strip with the same thickness but three times
the width, which shows less demagnetizing effect and, a result, a
smaller (in absolute value) initial slope of the magnetization.

Another feature observed in the interconnected case is the
observation of a kink (change in
the slope) in the magnetization curve, particularly for the cases of
large separation between strips. This effect is explained as
follows.
Since the magnetic moment is proportional to the area enclosed by the
loops, currents in the external strips contribute more to the
magnetization than those in the inner ones. So, when the external
strips become saturated, new current can only be induced in the
central
strip, having a lower contribution to the magnetization $M$, so that
the $M$ rate when $H_{\rm a}$ is increased is lower in magnitude; a
similar
effect has been predicted for rings in the critical state model
\cite{BrandtRing}. In a single strip or even in the case of an $x$
array with isolated filaments, this process is continuous, but not in
the present case of interconnected strips separated a horizontal
distance. 

\subsection{$xz$-matrices}

The magnetization of $xy$ matrices is again a combination of the
effects discussed above for horizontal arrays and in the
previous paper \cite{tapesa} for the vertical ones. In Fig.
\ref{f:Mhp2M} the initial magnetization curve
$M(H_{\rm a})$ for $xz$ matrices with the same vertical separation is
plotted, for a vertical separation
$h/a=0.2$, and several horizontal separations $d/a$. The curves are
qualitatively similar to those for $x$-arrays and the same values of
$d/a$, so that the discussion done for $x$-arrays is still valid. The
main difference between $x$-arrays and $xz$ matrices lies in both the
value of
the saturation field $H_s$, that is, the field which $M$ reaches its
saturation value, and the magnitude of the initial slope. For the
case of both isolated and interconnected $xz$ matrices, $H_s$ is
higher than for $x$-arrays, while the initial slope is lower. This is
due to the reduction of the demagnetizing effects owing to the
stacking in the $z$ direction \cite{tapesa,Mawatari,fabbricatore}.
Moreover, the mentioned differences of the $M(H_{\rm a})$ curve
between $x$
arrays and matrices would be qualitatively the same if we considered
an $x$-array with a larger filament thickness.
Detailed results of the magnetization of $xz$ matrices calculated by
our model will be presented elsewhere.

\section{AC losses}

\subsection{$x$-arrays}

In this section we study the imaginary part of the AC susceptibility,
$\chi''$, calculated form the magnetization loops obtained in section
IV, which can be easily related to the AC losses \cite{spectra}. In
Fig. \ref{f:XppAllA} we present calculated results for $\chi''$ as
function of the AC field amplitude $H_{\rm ac}$ for the same $x$
arrays discussed in the previous sections (with $b/a=0.1$ and
different separation distances $d/a=$2, 0.2, and 0.02). The two
different cases of interconnected and isolated strips are plot
together for comparison. Results show that the general trend is the
appearance of a peak in the $\chi''$ curve (and therefore 
a change of
slope of the AC losses). This peak, however, is wider for the case of
isolated strips (also shown in the figure), specially on the left
part of the peak. This effect has
been experimentally found in several works
\cite{GomAC_00,GomAC_02,SuenStack,Oomen}. Actually, the cause of the
disagreement between theoretical predictions and experiments in these
works is that they used models for single strips or disks, which
yielded narrow peaks. Our model allows for the explanation of this
effect.
Concerning hysteresis losses only, as we do in this work, the reason
for this widening of the peak is that the $M(H_{\rm a})$ curve
becomes
non linear at small applied field values because of the penetration
of magnetic flux not only in the outer surface regions of the tapes
but also in the channels between strips, where the field intensity is
enhanced. This deviation from linearity in the $M(H_{\rm a}$ curve
results in an increase of the
loss. 

We also find that decreasing the distance between strips results in a
higher or a smaller value of the peak, depending upon we are
considering the isolated or interconnected case, respectively. This
dependence on separation distance is only slight for the case of
isolated strips and much more evident for the interconnected ones.
These results can be understood from the magnetization curves of Fig.
\ref{f:MAllA}, in which we observe two important properties: the
initial slope of the magnetization curve for interconnected strips
increases (in absolute value) with increasing distance between
strips,
while it decreases for isolated strips, and, most important, for
interconnected strips, the saturation magnetization has very
different values for the different separations, while it remains
almost constant for isolated strips. All these effects have been
explained in section IV. Another characteristics observed in the two
upper curves of Fig. \ref{f:XppAllA} is a kink at a particular field
value, that is directly related to the presence of a similar kink
in the magnetization data shown in Fig. \ref{f:MAllA}. This kink was
already predicted for rings \cite{BrandtRing} and later
experimentally observed \cite{BrandtRing2}. Furthermore, experimental
evidence of a kink in actual superconducting tapes was shown for the
case of a Ag/Bi-2223
tape with the superconducting core shaped as a circular shell
\cite{Fukunaga} or two concentric elliptical shells \cite{GomAC_98}.

Another interesting result for the $\chi''$ calculations is shown in
Fig. \ref{f:Maw}, where we show the calculated results for $x$-arrays
of several strips with $b/a=0.01$ with a fixed separation distance of
$d/a=0.02$. Results are shown for arrays of 2, 3, 5, and 9 strips. We
consider the isolated strips case, in order to compare our
results with the analytical prediction for an infinite array of
Mawatari \cite{Mawatari}.
We also include the calculated result for a single strip with
$b/a=0.01$
as well as the same curve calculated from the analytical formulas for
thin
strips \cite{ThStrPRB}. The small difference between the two
latter results indicates that $b/a=0.01$ is already a satisfactory
value for using the thin strip approximation. On the other limit, we
check that the results for a large number of tapes tend to the
analytical results of Mawatari \cite{Mawatari}, although 9 is not a
sufficient number for approaching the limiting case (higher number of
tapes yield values closer to Mawatari's results; not shown for
clarity). The general
trend observed that the losses increase with the number of tapes is
due to the fact that the effect of the channels discussed above
increases for higher number of strips
\cite{fabbricatore,Farinon,Everett98}. 

\subsection{$xz$-matrices}

In Fig. \ref{f:Xpphp2M} we present the dependence of $\chi''$ upon
the AC applied field amplitude $H_{\rm ac}$ for $xz$ matrices with
$b/a=0.1$, $h/a=0.2$, and several values of $d/a$. It can be seen
that
the qualitative variations of the $\chi''(H_{\rm ac})$ curve when
considering isolated or interconnected strips is the same as for $x$
arrays, as well as the effect of changing $d/a$. However, for $xz$
matrices there is both a reduction of the peak in the
$\chi''(H_{\rm ac})$ curve and a shifting to higher $H_{\rm ac}$
values. These facts can be explained returning to the initial
magnetization
curves in Figs. \ref{f:MAllA} and \ref{f:Mhp2M}, where the initial
slope was lower for all $xz$ matrices and the saturation field was
higher. 
A detailed study of the AC losses from the $\chi''$ values, including
the real part of the susceptibility, $\chi'$, for $xz$ matrices
will be presented elsewhere.

\section{Conclusions}

We have presented a model that allow to study the response of a
horizontal
array of superconducting strips of finite thickness in a
perpendicular applied field. The different cases of isolated and
completely interconnected strips have been discussed separately. 
Current penetration results show that whereas in the interconnected
cases the filaments magnetically shield the whole internal volume of
the tape, in the case of isolated strips, the shielding is within
each of them. The latter effect in the isolated strip case creates
channels of field compression between
the strips, particularly when the separation distance between strips
is small. These channels govern the magnetic and AC losses properties
of the arrays of isolated tapes. Because of them, when decreasing
the horizontal distance between strips, the initial slope of the
magnetization curve increases (in absolute value), and,
correspondingly, there are larger AC losses. Moreover, the
experimentally found effect of a widening of the peak in the
imaginary part of the AC susceptibility can be explained by the same
effect. On the other hand, for the case of interconnected strips, 
the trend is the opposite: decreasing the horizontal distance between
strips the initial slope of the magnetization curve and the AC losses
are reduced. The effect governing these features are now the
demagnetizing effects: when strips are close to each other they
behave as a single tape with smaller aspect ratio and, therefore,
with smaller demagnetizing effects.

The magnetic properties of superconductor matrix arrays are a
composition of those for horizontal and vertical arrays, discussed
above and in the first paper of this series, respectively. A result
of practical importance is that AC losses are reduced when decreasing
the vertical separation between strips in the tape, because when
stacking strips in the vertical direction they behave as thicker
strips and therefore have less demagnetizing effects and less AC
losses.

In the present version, the model cannot be used to the study of the
case in which a transport current flows in the array in addition to
the applied magnetic field. This extension will be presented
elsewhere.

\section*{Acknowledgments}

We thank Fedor G\"om\"ory and Riccardo Tebano for comments.
We thank MCyT project BFM2000-0001, CIRIT project
1999SGR00340, and DURSI from Generalitat de Catalunya for financial
support. 

\appendix
\section{Field of full penetration}
\label{s:Hpen}

All previous results are calculated numerically. In this appendix we
provide some analytical calculations that may be useful in the
practice.

As explained in \cite{tapesa,Forkl,ThickStr}, the full penetration
field can be
calculated as minus the field created by the current distribution
${\bf H}_J$ in the last induced current point, where ${\bf
H}_J=H_J{\bf \hat{z}}$. Then, both the current distribution at the
penetration field and the last induced current point must be known to
calculate $H_{\rm pen}$.

For $x$-arrays and $xz$ matrices we differentiate again two cases
depending
on the way that the strips are connected at infinity: completely
interconnected strips and current isolated strips.

\subsection{Completely interconnected strips}

For this case, the volume current density at the penetration field is
${\bf J}=-J_c{\bf \hat{y}}$ for $x > 0$ and ${\bf J}=J_c{\bf
\hat{y}}$ for $x < 0$.

When both the number of strips in the $x$ axis $n_{f,x}$ and in the
$z$ axis $n_{f,z}$ are odd, the last induced current point ${\bf
r}_m$ is simply the center of the central strip.  Using the
Biot-Savart law to calculate $H_{J,z}({\rm r}=0)$, we obtain
\begin{eqnarray}
\label{HpnOO}
H_{\rm pen,matrix}(n_{f,x},n_{f,z})=H_{\rm
pen,stack}(n_{f,z})+&&\nonumber\\
{J_c\over 2\pi}\Bigg[
2\sum_{i=1}^{n_{f,x}-1\over 2}F_2((2a+d)i,0,a,b)+&&\nonumber\\
4\sum_{i=1}^{n_{f,x}-1\over 2}\sum_{j=1}^{n_{f,z}-1\over 2}
F_2((2a+d)i,(2b+h)j,a,b)\Bigg],&& 
\end{eqnarray}
where $H_{\rm pen,stack}$ is the penetration field for a $z$-stack
\cite{tapesa} and the function $F_2(u,v,t,d)$ is defined as
\begin{eqnarray}
\label{F2def}
&& F_2(u,v,t,d)=(u-t)\bigg[ 
\arctan\left({{v-d}\over{u-t}}\right)-\nonumber\\
&&\arctan\left({{v+d}\over{u-t}}\right)\bigg]+
(u+t)\bigg[ \arctan\left({{v+d}\over{u+t}}\right)-\nonumber\\
&&\arctan\left({{v-d}\over{u+t}}\right)\bigg]+ 
{(u-d)\over 2}\ln\left[{{(u-t)^2+(v-d)^2\over
(u+t)^2+(v-d)^2}}\right]+ \nonumber\\
&&{(u+d)\over 2} \ln\left[{{(u+t)^2+(v+d)^2\over
(u-t)^2+(v+d)^2}}\right].
\end{eqnarray}
The penetration field for an $x$-array with an odd number of strips
is the same as in Eq. (\ref{HpnOO}) but removing the term with the
double sum and taking $n_{f,z}=1$.

When either $n_{f,x}$ or $n_{f,z}$ are even, the last induced current
point is not easy to be determined. In those strips that current
returns
through the same filament, the total magnetic field increases
monotonically from the edges of the strip to the current profile.
When
$n_{f,x}$ is odd and $n_{f,z}$ is even the last strips to be fully
penetrated are those in the central column and in the inner rows.
Then, the last induced current point ${\bf r}_m$, where $H_{J,z}({\bf
r}_m)=-H_{\rm pen}$, is on the $z$ axis and can be determined as the
point where $H_{J,z}$ is maximum in absolute value. When $n_{f,x}$ is
even, we have found no way to analytically calculate ${\bf r}_m$ and
$H_{\rm pen}$.

\subsection{Current Isolated Strips}

As discussed in Sec. IIIB, the current interface at the penetration
field
is almost a vertical straight line at the center of the strip. We
have found that this
approximation is reasonable even for strips with a ratio
$b/a$ as large as $b/a=1$. 

When $n_{f,z}$ is odd, the last penetrated current point is at the
center of the strips belonging to the central row and the most
external columns. This is so because external rows shield inner ones
and external columns increase the field on the inner ones. Then,
using the Biot-Savart law and assuming straight current
interfaces, the penetration field for a $xz$-matrix with odd
$n_{f,z}$ is
\begin{eqnarray}
\label{HpnOpm}
H_{\rm pen,matrix}(n_{f,x},n_{f,z})=&&\nonumber\\
{-J_c\over 2\pi}\Bigg[
\sum_{i=0}^{n_{f,x}-1}F_3((2a+d)i,0,a,b)+&&\nonumber\\
2\sum_{i=0}^{n_{f,x}-1}\sum_{j=1}^{n_{f,z}-1\over 2}
F_3((2a+d)i,(2b+h)j,a,b)\Bigg],&&
\end{eqnarray}
where the function $F_3(u,v,t,d)$ is defined as
$F_3(u,v,t,d)=F_2(u-{t/2},v,{t/2},d)-F_2(u+{t/2},v,{t/2},d)$.
Notice that Eq. (\ref{HpnOpm}) is valid when $n_{f,x}$ is either odd
or even, while Eq.(\ref{HpnOO}) is only valid 
for an odd $n_{f,x}$. The penetration field for an $x$-array is the
same as described in Eq. (\ref{HpnOpm}) but removing the term with
the double sum.

\begin{figure}
\caption{Sketch of the array of superconducting tapes. A $xz$-matrix
is drawn, although all the parameters described are also valid for
$x$-arrays. The $y$ axis is perpendicular to the plane and it is
oriented inwards.}\label{f:SkGn}
\end{figure}

\begin{figure}
\caption{Sketch of the real closed current loops (solid thick lines)
and those used in the simulation (dashed thick lines).
The case of an $x$-array with two strips is drawn for simplicity.
Four current elements are represented as elongated thin rectangular
prisms where a single straight current flows following the $y$
axis.}\label{f:SkIs}
\end{figure}

\begin{figure}
\caption{Total magnetic flux lines and current profiles for
interconnected $x$-arrays at an applied field of $H_{\rm a}=0.2H_{\rm
pen}$,
being $H_{\rm pen}$ the complete penetration field for the whole
$x$-array. The strips in the arrays have an aspect ratio $b/a=0.1$
and
the distances between strips are: 
(a) $d/a=$0.02, (b) $d/a=$0.2, and (c) $d/a=$2. The horizontal scale
has been contracted for
clarity, while the vertical scale is the same for all
figures.}\label{f:FfInA}
\end{figure}

\begin{figure}
\caption{Total (left) and self (right) magnetic field lines and
current profiles for interconnected $xz$ matrices at an applied field
of $H_{\rm a}=0.2H_{\rm pen}$. For the strips $b/a=0.1$ and
$d/a=h/a$=0.02 (a,b), 0.2 (c,d), and 2 (e,f). Vertical and horizontal
scales are rescaled for clarity.}\label{f:FfInM}
\end{figure}

\begin{figure}
\caption{The same as Fig. \ref{f:FfInA} but with isolated strips. The
applied field is $H_{\rm a}=0.1H_{\rm pen}$ and the strips have
dimensions
$b/a=0.1$ spaced a distance: (a) $d/a=$0.02, (b) $d/a=$0.2, and (c)
$d/a=$2. The horizontal scale has been contracted for
clarity.}\label{f:FfIsA}
\end{figure}

\begin{figure}
\caption{Current profiles for $x$-arrays with $b/a=0.1$ and $d/a=0.2$
for (a) interconnected strips and (b) isolated strips. The
vertical axis has been expanded for clarity. The applied field values
corresponding to each current profile are $H_{\rm a}=$0.1, 0.2, 0.4,
0.6, 0.8 and
1 in units of the penetration field $H_{\rm pen}$ for each case.
}\label{f:Cpdp2}
\end{figure}

\begin{figure}
\caption{Total (left) and self (right) magnetic field lines and
current profiles for isolated $xz$ matrices at an applied field
$H_{\rm a}=0.1H_{\rm pen}$. For the strips $b/a=0.1$ and $d/a=h/a$=
0.02 (a,b), and 2 (c,d).}\label{f:FfIsM}
\end{figure}

\begin{figure}
\caption{Initial magnetization curves $M(H_{\rm a})$ for $x$-arrays
with three strips with $b/a=0.1$ and several strip separations
$d/a$ for the cases of (a) isolated strips and (b) interconnected
strips. For graph (a) solid lines correspond to $x$-arrays with
$d/a=$2, 0.2, and 0.02 from top to bottom, while the dashed line
represents
$M(H_{\rm a})$ for a single strip with $b/a=0.1$. For graph
(b) solid lines correspond to $x$-arrays with $d/a=$0.02, 0.2, and 2
from top to
bottom and the dashed line is for a single strip with halfwidth
$a'=3a$
and $b=0.1$.}\label{f:MAllA}
\end{figure}

\begin{figure}
\caption{Initial magnetization curves $M(H_{\rm a})$ for
$xz$-matrices with $3\times 3$ strips of dimensions $b/a=0.1$, for
$h/a=0.2$ and
several values of $d/a$ for the cases of (a) isolated strips and (b)
interconnected strips. For graph (a) curves correspond to
$d/a=$2, 0.2, and 0.02 from top to bottom. For graph (b) curves
correspond
to  $d/a=$0.02, 0.2, and 2 from top to bottom.}\label{f:Mhp2M}
\end{figure}

\begin{figure}
\caption{Imaginary AC susceptibility $\chi''$ as a function of the AC
applied field amplitude $H_{\rm ac}$ 
corresponding to the $M(H_{\rm a})$ curves showed in Fig.
\ref{f:MAllA} for
$x$-arrays. The strips dimensions are $b/a=0.1$. Solid lines
are for the case of interconnected strips for $d/a=$2, 0.2, and 0.02
from
top to bottom, while dashed lines are for isolated strips with
$d/a=$0.02, 0.2, and 2 from top to bottom.}\label{f:XppAllA}
\end{figure}

\begin{figure}
\caption{Imaginary AC susceptibility $\chi''$ as a function of
$H_{\rm ac}$ for $x$-arrays with several numbers of strips $n_f$
(solid lines), corresponding to $n_f=9,5,3,2$ from top to bottom,
compared to a single strip (dashed line) and the analytical limits
for a single thin strip (lower dotted line) and an infinite $x$
array of thin strips (upper dotted line). The strips dimensions are
$b/a=0.01$.}\label{f:Maw}
\end{figure}

\begin{figure}
\caption{Imaginary AC susceptibility $\chi''$ as a function of
$H_{\rm ac}$ corresponding to the $M(H_{\rm a})$ curves showed in
Fig.
\ref{f:Mhp2M} for $xz$ matrices. The strips dimensions are $b/a=0.1$
and the vertical separation is fixed, being $h/a=0.2$. Solid
lines are for the case of interconnected strips for $d/a=$2, 0.2, and
0.02
from top to bottom, while dashed lines are for isolated strips with
$d/a=$0.02, 0.2, and 2 from top to bottom.}\label{f:Xpphp2M}
\end{figure}

\end{document}